\documentclass{aa}  

\usepackage{graphicx}
\usepackage{txfonts}
\usepackage{graphicx}
\usepackage{amsmath}
\usepackage{amssymb}
\usepackage{amsfonts}
\usepackage[dvipsnames]{xcolor}
\usepackage[linktoc=none]{hyperref}
\usepackage[utf8]{inputenc}
\usepackage{multirow}
\usepackage{natbib}
\bibpunct{(}{)}{;}{a}{}{,}

\hypersetup{colorlinks,linkcolor={blue},citecolor={teal},urlcolor={violet}}

\begin{document}

   \title{Observable Signatures of Cosmic Rays Transport in Starburst Galaxies on Gamma-ray and Neutrino Observations}
   \newcommand{\shorttitle}{Observable Signatures of CRs Transport in SBGs}
   

   \author{A. Ambrosone,
          \inst{1,2}
         M. Chianese, \inst{1,2}
         D.F.G. Fiorillo, \inst{1,2,3}
         A. Marinelli\inst{2} 
         \and
         G. Miele\inst{1,2,4}
          }

   \institute{Dipartimento di Fisica ``Ettore Pancini'', Università degli studi di Napoli ``Federico II'', Complesso Univ. Monte S. Angelo, I-80126 Napoli, Italy
         \and
             INFN - Sezione di Napoli, Complesso Univ. Monte S. Angelo, I-80126 Napoli, Italy
            \and
            Niels Bohr International Academy, Niels Bohr Institute, University of Copenhagen, Copenhagen, Denmark
            \and
            Scuola Superiore Meridionale, Università degli studi di Napoli ``Federico II'', Largo San Marcellino 10, 80138 Napoli, Italy
             }
             
   \offprints{A. Ambrosone,\\ \email{aambrosone@na.infn.it}}

   \date{}

 
  \abstract
  {The gamma-ray emission from Starburst and Starforming Galaxies (SBGs and SFGs) strongly suggest a correlation between star-forming activity and  gamma-ray luminosity. However, the very nature of cosmic-ray (CR) transport and the degree of their confinement within SBG cores are still open questions.}
  {We aim at probing the imprints left by CR transport on gamma-ray and neutrino observations of point-like SFGs and SBGs, looking into quantitative ways to discriminate among different transport models. Moreover, following the reported scenarios, we quantitatively assess the SBGs and SFGs contribution to the Extra-galactic Gamma-Ray Background (EGB data) and the IceCube diffuse observations (HESE data).}
  {We analyse the 10-year Fermi-LAT spectral energy distributions of 13 nearby galaxies with two different CR transport models, taking into account the corresponding IR and UV observations. We generate mock gamma-ray data to simulate the CTA performance in detecting these sources. In the way, we propose a test to discriminate between the two CR models, quantifying the statistical confidence at which one model can be preferred over the other.}
  {We point out that current data already give a slight preference to CR models which are dominated by advection in their nucleus. Moreover, we show that CTA will allow us to firmly disfavour models dominated by diffusion over self-induced turbulence, compared to advection-dominated models, with Bayes factors which can be as large as $10^7$ for some of the SBGs. Finally, we estimate the diffuse gamma-ray and neutrino fluxes of SFGs and SBGs, showing that they can explain $25\%$ of the diffuse HESE data, while remaining consistent with gamma-ray limits on non-blazar sources.}
  {Our results point out that future gamma-ray telescopes such as CTA and SWGO will constrain the high-energy emission from nearby galaxies, thus providing remarkable insights into the CR transport within SFGs and SBGs. In particular, they will potentially confirm these sources as efficient high-energy neutrino factories contributing to the diffuse neutrino flux.}

   
    \titlerunning{\shorttitle}
    \authorrunning{Ambrosone et al.}
   \maketitle
%

\section{Introduction \label{sec:intro}}

Starburst and Starforming Galaxies (SBGs and SFGs) are astrophysical sources with a compact central region, called the starburst nucleus (SBN), where the star-forming activity is mostly concentrated. They are endowed with a large gas density  $(n_{\rm ISM} \sim 100-1000 \ \text{cm}^{-3})$, an enhanced supernova explosion rate  $R_{\rm SN} \approx 0.1 - 1 \ \text{yr}^{-1}$ and magnetic fields of the order of  $10^{2} \ \ \mu \text{G}$ \citep{Thompson:2006is}. The measurements by  the Fermi-LAT~\citep{Ajello:2020zna}, VERITAS~\citep{2009Natur.462..770V} and H.E.S.S.~\citep{2018A&A...617A..73H} collaborations have shown their capability to produce gamma-rays \citep{Ajello:2020zna} with a luminosity correlated to their star formation rate (SFR). This suggests that the non-thermal emission of these galaxies can be strictly dependent on their star-forming activity~\citep{Ackermann_2012, Ajello:2020zna,Kornecki:2020riv,Kornecki:2021xiy}. Indeed, supernova remnants (SNRs) shocks are supposed to accelerate protons up to PeV energies~\citep{Sveshnikova:2003sa,Murase:2013rfa,Tamborra:2014xia}. Furthermore, if high-energy cosmic-rays (CRs) are confined within the SBN by diffusion on magnetic inhomogeneities for long times, they have an enhanced chance to copiously produce gamma-rays and neutrinos through hadronic collisions. In fact, SBGs and SFGs are often considered to be astrophysical calorimeters or reservoirs, being able to confine high-energy cosmic-rays within their very core. The calorimetric condition requires the timescale for energy loss of CRs to be shorter than their escape time from the SBN. However, although electron confinement is nowadays an established result~\citep{Peretti:2018tmo,Roth:2021lvk}, the degree of high-energy proton calorimetry is still under debate. This is because it depends on the interplay between different phenomena like diffusion, wind advections and energy losses, which are difficult to reliably model. Different CR transport scenarios have been used to describe the observed gamma-ray emission from nearby SFGs and SBGs, trying to disentangle a calorimetric behaviour from a  diffusion-dominated regime ($\tau_{\rm diff}\ll \tau_{\rm loss},\tau_{\rm adv}$)~\citep{Peretti:2018tmo,Krumholz:2019uom,Ha:2020nty,Muller:2020vdm,Shimono:2021wvp,Xiang:2021amz,Owen:2021evt,Owen:2021qul,Werhahn:2021jvy,Werhahn:2021bal,Sudoh:2018ana,Kornecki:2020riv,Kornecki:2021xiy}.
The major uncertainty factor is due to distinctive timescale modelling. Different models suggest Kolmogorov-like diffusion from magnetic field turbulence~\citep{Yoast-Hull:2013wwa,Peretti:2018tmo,Peretti:2019vsj} as well as  self-generated diffusion generated by the streaming instability~\citep{Krumholz:2019uom,Roth:2021lvk}. Uncertainty comes also from the fraction of hot ionized gas (the target for proton-proton interactions) which can vary from $10^{-4}$ to $10^{-1}$~\citep{Krumholz:2019uom,Peretti:2018tmo}. This, in turn, leads to a different role for the advection, which predominately affect the hot ionized gas in SBNs~\citep{Krumholz:2019uom}. Furthermore, a greater fraction of neutral gas would reduce Kolmogorov-like turbulence, supporting streaming instability as source of magnetic inhomogeneities.

In this framework, current measurements make it challenging to quantitatively discriminate between different scenarios, leading to a significant uncertainty in the gamma-ray and neutrino fluxes expectation from SBGs. A major limitation comes from the fact that only 13 SFGs are observed through gamma rays and just a few of them are observed at energies greater than $10 \ \text{GeV}$~\citep{Ajello:2020zna,Kornecki:2020riv}. This points out the importance for future gamma-ray experiments~\citep{CTAConsortium:2018tzg,Albert:2019afb,Hinton:2021um} as well as possible neutrino point-like observations~\citep{KM3Net:2016zxf,Aiello:2018usb,Aartsen:2019fau,Ambrosone:2021aaw} to better constrain the characteristics and the CR transport scenario for the nuclei of SBGs.

In this paper, we assess the signatures of different cosmic-ray transport mechanisms on point-like and diffuse gamma-ray and neutrino contribution of SFGs and SBGs, and exploit them to understand which model can explain better the data. In particular, we consider both an advective model adopted, e.g., by~\cite{Peretti:2019vsj}, and the model provided by~\cite{Krumholz:2019uom}, which considers CR transport in the nucleus dominated by diffusion on self-induced turbulence. Following the analysis described in~\citep{Ambrosone:2021aaw}, we compare the gamma-ray predictions of the two CR transport models with 10-year Fermi-LAT data \citep{Ajello:2020zna} as well as H.E.S.S.~\citep{2018A&A...617A..73H} and VERITAS~\citep{2009Natur.462..770V} collaborations data for 13 known SBGs (VERITAS data are available only for M82 and H.E.S.S. data are only available for NGC 253). As in~\citep{Ambrosone:2021aaw}, for each source we require the star formation rate (SFR) to be in agreement with the one derived from infrared and ultraviolet observations~\citep{Kornecki:2020riv}. We show that current data are slightly better accommodated by the model developed by~\cite{Peretti:2018tmo}, suggesting advection to be important for high-energy CR transport in SBNs. Then, we investigate the ability of future gamma-ray telescopes sensitive to $\mathcal{O}(10~{\rm TeV})$ photons, such as CTA~\citep{CTAConsortium:2018tzg} and SWGO~\citep{Albert:2019afb,Hinton:2021um}, to further discriminate between the different CR transport models. We perform a forecast analysis by means of mock data generation for the most promising sources using the public information of the CTA telescope. In particular, we determine the statistical confidence at which the diffuse-dominated model~\citep{Krumholz:2019uom} might be excluded in the near future in favour of the advection-dominated one~\citep{Peretti:2019vsj}.\footnote{By diffusion or advection dominated models, we mean that diffusion and advection are respectively the main escape processes for those particular models.} We report the $p$-values and the Bayesian factors obtained in case of Frequentist and Bayesian approaches, respectively. We hence emphasize the importance of future gamma-ray telescopes for constraining the SFGs and SBGs emission properties.

Finally, we calculate the diffuse neutrino and gamma-ray fluxes from the whole populations of SFGs and SBGs making use of the approach put forward by~\cite{Ambrosone:2020evo}. We set the properties of all the sources, e.g. the distribution of the spectral indexes, to be consistent with the ones previously inferred by the local point-like observations. Interestingly, such a comprehensive data-driven scenario may explain $25\%$ of the HESE neutrino flux~\citep{IceCube:2020wum} when a cut-off energy of protons of the order of $10 \ \text{PeV}$, as well as the $33\%$ of the Extra-galactic Gamma-ray Background (EBG) above 50~GeV~\citep{TheFermi-LAT:2015ykq,Lisanti:2016jub,Bechtol:2015uqb,Yoshida:2020div}. Hence, it is in agreement with the independent multi-component fit provided by~\cite{Ambrosone:2020evo}, and it is consistent within $1\sigma$ with the gamma-ray limits on non-blazar sources~\citep{TheFermi-LAT:2015ykq,Lisanti:2016jub,Bechtol:2015uqb,Yoshida:2020div}. This result becomes crucial when compared to the one presented by~\cite{Roth:2021lvk}, since it shows that the use of a different CR transport model for the SBG class allows us to explain a greater portion of diffuse high-energy neutrino measurements without evading all the existing EGB bounds on non-blazar sources.

The paper is organized as follows. In Sec.~\ref{sec:cosmic}, we describe in details the two CR models we utilize for our analyses and how we calculate gamma-ray and neutrino spectra. In Sec.~\ref{sec:gamma}, we describe the analysis on current gamma-ray data while in Sec.~\ref{sec:forecast} we discuss the forecast analysis for the CTA telescope. In Sec.~\ref{sec:diffuse}, we calculate the diffuse gamma-ray and neutrino fluxes. Finally, in Sec.~\ref{sec:conclusion}, we draw our conclusions.

\section{Cosmic-rays transport and non-thermal emissions in a SBG\label{sec:cosmic}}

In this section we model the transport of cosmic-rays, with particular emphasis on the differences between the two approaches of~\cite{Peretti:2018tmo} and~\cite{Krumholz:2019uom} (hereafter, model A and model B, respectively). The phenomenology arising from these differences in the gamma-ray and neutrino fluxes is the fundamental topic of this paper. Since the starburst activity lasts for long time ($\sim$\,$ 10^{8} \ \text{yr}$ \citep{Peretti:2021yhc}), a steady state is reached between the cosmic-ray cooling, transport, and injection phenomena. In particular, the distribution $F_{p}(E)$ of high-energy protons with energy $E$ can be written as 
\begin{equation}\label{proton}
    F_{p} (E) = Q(E)\cdot \tau_{\text{life}}(E) = Q(E)\cdot \tau_{\text{loss}}(E)\cdot F_{\text{cal}}(E)
\end{equation}
where $Q(E)$ is the injection rate of protons,  $\tau_{\text{life}}$ is the lifetime of protons inside the SBN, and $\tau_{\text{loss}}$ is the losses timescale. In both models, the energy losses of protons are defined by the the processes of p-p interactions, ionization and Coulomb interaction \cite{Peretti:2018tmo,2002cra..book.....S}. However, the effects due to  Coulomb and ionization interactions are negligible for $E > 1\, \text{GeV}$. Finally, $F_{\text{cal}} (E)$ is the so-called calorimetric fraction, which represents the fraction of high-energy protons which effectively lose their energy via proton-proton collisions inside the nucleus, thereby producing gamma-rays and neutrinos. If SBGs were efficient proton calorimeters at all energy ranges, we would have $F_{\text{cal}} (E) = 1$. Models A and B induce different calorimetric fractions because of different assumptions on CR transport and geometry. We are going to describe these separately in the following two subsections.

\subsection{Model A for cosmic-ray transport}

Model A (see~\cite{Peretti:2018tmo} for more details) considers the SBN as a spherical compact region and adopts a leaky-box model to compute the proton distribution. In this case, the calorimetric fraction takes the following form
\begin{equation}\label{fractionperetti}
    F_{\text{cal}} (E) = \frac{\tau_{\text{eff}}(E)}{\tau_{\text{eff}}(E) + 1}
\end{equation}
where $\tau_{\text{eff}} (E)$ is the dimensionless effective optical depth for protons, which represents the effective depth of the material through which a CR must pass on its way out of the SBN. It is given by the ratio between the escape time $\tau_\mathrm{esc}$ for protons and the energy losses timescale $\tau_\mathrm{loss}$. The escape time is
\begin{equation}
    \tau_\mathrm{esc} = \left(\frac{1}{\tau_\mathrm{adv}} + \frac{1}{\tau_\mathrm{diff}}\right)^{-1},
\end{equation}
where both the advection ($\tau_\text{adv}$) and the diffusion ($\tau_\text{diff}$) timescales do not depend on the position in the nucleus. The former is $\tau_{\text{adv}} = R_{\rm SBN}/v_{\text{wind}}$ where $R_{\rm SBN} = 200\,{\rm pc}$ and $v_{\text{wind}} = 500\,{\rm km/s}$ are the radius of the SBN and the wind velocity, respectively (assumed equal for all the galaxies). The latter is $\tau_{\text{diff}} \propto E^{-1/3}$ according to a Kolmogorov-like scenario with an energy density of the magnetic field $F(k) \propto k^{-2/3}$ and a regime of strong turbulence inside the SBN~\citep{Peretti:2018tmo,Peretti:2019vsj}. Since the SBN extension is typically hundreds of parsecs, we can neglect a possible radial dependence of the diffusion coefficient as it is instead observed in our Galaxy~\citep{Erlykin:2012dp,Gaggero:2015xza,Gaggero:2014xla}. We adopt an average value of $200~\mu{\rm G}$ for the magnetic field~\citep{Thompson:2006is}. Under reasonable values for the parameters, model A predicts a high degree of calorimetric regime with a CR lifetime dominated by p-p loss timescale. The Kolgomorov-like diffusion has only a marginal impact on CR transport, which is predominantly affected by p-p losses and advection as a main escape mechanism. As we will see, this is the main difference with the model B.

\subsection{Model B for cosmic-ray transport}

Model B (see~\cite{Krumholz:2019uom} for more details) considers the nucleus as a cylinder with a width of the order of $10^{2} \ \text{pc}$ and neglects any advective phenomenon. In this case, the calorimetric fraction is given by
\begin{equation}\label{calorimetrickrumholz}
    F_{\text{cal}} = 1 - \left[\,_0F_1\left(\frac{1}{5}, \frac{16 }{25}\tau_{\text{eff}}\right) + \frac{3 \tau_{\text{eff}}}{4 M_{A}^{3}} \,_0F_1\left(\frac{9}{5}, \frac{16 }{25}\tau_{\text{eff}}\right)\right]^{-1}
\end{equation}
with $M_{A}\simeq 2$. Eq.s~\eqref{fractionperetti} and~\eqref{calorimetrickrumholz} generally give different predictions for the calorimetric fraction. In particular, under the same $\tau_{\rm eff}$, Eq. ~\eqref{calorimetrickrumholz} predicts a higher calorimetric fraction than Eq.~\eqref{fractionperetti}. This is due to the difference in the SBN geometry of the two models. Furthermore, differently from model A, in model B CRs do not scatter off the strong large-scale turbulence of the magnetic field, but instead stream along field lines at a rate determined by the competition between streaming instability and ion-neutral damping, leading to transport via a process of field line random walk~\citep{Krumholz:2019uom}. In this case, the diffusion stems from the interaction with the Alfven waves that CRs themselves generate via the streaming instability. We estimate the diffusive timescale following~\cite{Krumholz:2019uom}. Firstly, we consider the velocity of the Alfven waves as
\begin{equation}
    V_{\rm al} = \frac{\sigma_{g}}{\sqrt 2\,\chi^{1/2} \, M_{A}}
\end{equation}
where $\chi = 10^{-4}$ is the ionisation fraction and $\sigma_{g}$ is the dispersion velocity for which we use the same scaling relation with the star formation rate reported by~\cite{Roth:2021lvk}). Then, we calculate the streaming velocity of CRs with energy $E$ as
\begin{align}
V_{\rm st} = & \, {\rm min}\bigg[c,\, V_{\rm al}\,\bigg(1+ 2.3 \times 10^{-3} \ c_{3}^{-1} \bigg(\frac{E}{m_p}\bigg)^{\Gamma -1}\bigg(\frac{n_{\rm ISM}}{10^3 \ \text{cm}^{-3}}\bigg)^{3/2} \nonumber \\ 
& \qquad \qquad \qquad \qquad \qquad \times \bigg(\frac{\chi}{10^{-4}}\bigg)  \bigg(\frac{\sigma_{g}/\sqrt2}{10 \ \text{km}\ \text{s}^{-1}} \bigg)^{-1}\bigg) \bigg]
\end{align}
where $m_p$ and $c$ are the proton mass and the speed of light, respectively. The factor $c_3\approx 1$ is the ratio between the number density of CRs in the middle plane in the SBG and a thousand times the number density of CRs expected in the Milky Way near the Solar circle.\footnote{We have verified that a slightly different value for $c_3$ does not significantly affect our results}. Moreover, $n_{\rm ISM}$ denotes the  interstellar medium density (for SBGs $n_{\rm ISM} \sim 10^{2-3}\ \text{cm}^{-3}$) and $\Gamma + 2$ is the spectral index of the proton injection spectrum. It is worth mentioning that the diffusion mechanism breaks down at high energies, where the streaming velocity becomes equal to the speed of light. In this regime, the protons start to free-stream out of the SBN. From the streaming velocity the diffusive coefficient can be calculated as
\begin{equation}
    D = V_{\rm st} \,\cdot L_{A}
\end{equation}
where $L_{A} = h/{\rm min}[1,M_{A}^{3}]$ is the turbulence length scale with $h$ being the height of the galactic disk. For the sake of simplicity, we consider $h = 73\, {\rm pc}$ for all the galaxies. Finally, the diffusion timescale is $\tau_{\rm diff} = h^2/D$.

\subsection{Non-thermal emissions}

In both models A and B, in order to calculate the gamma-ray and neutrino spectra, we assume that protons are injected with a power-law spectrum in momentum space with spectral index $\Gamma+2$. The proton spectrum is directly proportional to the star formation rate $\dot{M}_*$ and normalized by requiring that each supernova releases into protons $10\%$ of its total explosion kinetic energy ($\sim 10^{51}~\mathrm{erg}$). Moreover, it is characterised by an exponential cutoff at $10~\mathrm{PeV}$, in agreement with the combined fit of IceCube and Fermi-LAT diffuse data~\citep{Ambrosone:2020evo}. We take also into account the injection of primary electrons featuring a power-law spectrum in momentum space with spectral index $\Gamma+2$, a normalization   equal to 1/50 of the one of the protons, and a Gaussian cutoff at $10~\mathrm{TeV}$ similar to what is inferred for our Galaxy~\citep{Torres:2004ui,Peretti:2018tmo}. The distribution of electrons can be computed in a similar way as Eq.~\eqref{proton} where we take $F_{\rm cal} = 1$~\citep{Peretti:2018tmo,Peretti:2019vsj,Roth:2021lvk}.

Neutrinos are emitted through the decay of charged pions ($\pi \to \mu \,\nu_\mu$, $\mu \to e\,\nu_e\,\nu_\mu$) that are produced in hadronic interactions of the injected protons with the interstellar gas with density $n_{\rm ISM}$. We determine the neutrino production rate by assuming that pions always carry 17\% of their parent proton energy~\citep{Kelner:2006tc}. For each galaxy, the density $n_\mathrm{ISM}$ is obtained by means of the same scaling relation with the SFR reported by~\cite{Ambrosone:2021aaw}, in agreement with the  Kennicutt relation~\citep{Kennicutt:1998zb,Kennicutt:2012ea,Kennicutt_2021}. Such a relation connects the surface density of SFR, $\Sigma_{\rm SFR} $, and the gas surface density, $\Sigma_{\text{gas}}$. In particular, we have
\begin{equation}\label{eq:kennicutscaled}
    n_{\rm ISM} = 175 \left(\frac{\dot{M}_*}{5~{\rm M_\odot\, yr^{-1}}}\right)^{2/3}\,{\rm cm^{-3}} \,.
\end{equation}

Gamma-rays are principally emitted in hadronic processes through neutral pion decays ($\pi^0\to \gamma\gamma$). Nonetheless, we also take into account the gamma-ray emission from bremsstrahlung and Inverse Compton scatterings of primary electrons as well as secondary ones (see~\cite{Peretti:2018tmo} and \cite{Ambrosone:2021aaw} for details). The gamma-ray emission from Inverse Compton scatterings depends on the density of background photons acting as a target. We consider all sources to have a similar spectral shape for the background photons equal to the M82 best-fit background spectrum reported by~\cite{Peretti:2018tmo}. The normalization for the different sources is self-consistently determined by the radiation energy density $U_{\text{rad}}$ of each source. For such a quantity, we assume a direct proportionality to the SFR, which is expected to be tightly related to the infrared (IR) luminosity~\citep{Kennicutt:1998zb,Inoue:2000hm,Hirashita:2003su,2011PASJ...63.1207Y,Kennicutt:2012ea,Kennicutt_2021}. 
\begin{equation}\label{eq:infrared}
    U_\mathrm{rad} = 2500 \left(\frac{\dot{M}_*}{5 \, {\rm M_\odot\, yr^{-1}}}\right)\,{\rm eV \,cm^{-3}} \,,
\end{equation}
We emphasize that both the hadronic and secondary leptonic components are dictated by the star formation rate and the calorimetric fraction, namely by the CR transport mechanisms assumed in the SBNs. Finally, we account for internal and external gamma-ray absorption due to pair-production processes with background photons. For the latter, we consider as a target CMB photons as well as the extragalactic background light model reported in~\cite{Franceschini:2017iwq}.

\section{Imprints of the cosmic-ray transport on current gamma-rays data\label{sec:gamma}}

The models A and B, outlined in the previous section, are based on different assumptions on CRs transport, and this gives rise to differences in the gamma-ray and neutrino spectra. This is highly significant because for nearby SBGs we have available both gamma-ray data and estimates of their SFR through IR and UV data. Therefore, the comparison of the theoretical predicted gamma-ray spectra with data allows us to scrutinise whether there are observable features characterising the CR transport inside SBGs. We here discuss such a comparison for models A and B with actual data.

We study the gamma-ray spectral energy distributions (SEDs) of 13 galaxies observed by Fermi-LAT in 10 years of observations \citep{Ajello:2020zna}.  For M82 and NGC~253 we also use the data provided by VERITAS~\citep{2009Natur.462..770V} and H.E.S.S.~\citep{2018A&A...617A..73H}, respectively. For each galaxy, as performed by Ref. \citep{Ambrosone:2021aaw}, we determine the most-likely values for the two free parameters of the two models: the star formation rate $\dot{M}_*$ and the spectral index $\Gamma$ of injected protons and electrons. We adopt a Bayesian approach, using as a posterior distribution
\begin{equation}\label{eq:post}
    p(\dot{M}_*,\,\Gamma|{\rm SED}) \propto \mathcal{L}({\rm SED} | \dot{M}_*,\,\Gamma)\, p(\dot{M}_*) \, p(\Gamma)
\end{equation}
with a Gaussian likelihood function
\begin{equation}\label{eq:like}
    \mathcal{L}({\rm SED} | \dot{M}_*,\,\Gamma) = \exp \left[-\frac{1}{2}\sum_{i}\left(\frac{\text{SED}_{i}-E^2_{i}\Phi_{\gamma}(E_i| \dot{M}_*,\,\Gamma)}{\sigma_{i}}\right)^2 \right]
\end{equation}
Here, $\text{SED}_{i}$ are the measured data, where $i$ runs over the energy bins centered around the energy $E_{i}$, and $\sigma_{i}$ are the observational uncertainties. We compare the data with the gamma-ray flux $\Phi_{\gamma}(E_i| \dot{M}_*,\,\Gamma)$ predicted by the models considered in each energy bin. We assume the source distances given in \cite{Kornecki:2020riv}. Following the analysis of Ref. \citep{Ambrosone:2021aaw}, for all the galaxies we consider the same uniform prior $p(\Gamma)$ on the spectral index in the range 1.0--3.0. For $\dot{M}_*$, we account for prior information driven by the current measurements of the SFR, assuming a prior distribution $p(\dot{M}_*)$ uniform in an interval of a factor of 3 from the SFR values given in~\cite{Kornecki:2020riv}. This choice is only representative of the wide variety of SFR estimates present in the literature~\citep{Groves_2008,Bolatto_2011,For_2012,10.1093/mnras/stv2951,Yoast-Hull:2017wwl,Peretti:2019vsj,Ajello:2020zna}.
\begin{table}[t!]
\caption{\label{tab:data}Results of the Bayesian inference with current gamma-ray data for models A and B. Reported are the most-likely values for the star formation rate $\dot{M}_*$ in ${\rm M_\odot\, yr^{-1}}$ and the spectral index $\Gamma$, along with the reduced chi-square $\chi^2$. The results for model A have already been reported by~\cite{Ambrosone:2021aaw}.}
\centering
\begin{tabular}{l |c c| c c}
\multirow{2}{*}{\small{Source}}& \multicolumn{2}{c|}{\small{Model A}} & \multicolumn{2}{c}{\small{Model B}} \\
 & \small{$(\dot{M}_*,\,\Gamma)$} & \small{$\chi^2 / \mathrm{dof}$} & \small{$(\dot{M}_*,\,\Gamma)$} & \small{$\chi^2 / \mathrm{dof}$} \\
\hline                    
\small{M82} & \small{(4.5, 2.30)} & \small{1.24} & \small{(3.0, 2.30)} & \small{1.88} \\ 
\hline
\small{NGC 253} & \small{(3.3, 2.30)} & \small{1.32} & \small{(2.1, 2.20)} & \small{1.95} \\
\hline
\small{ARP 220} & \small{(740, 2.66)} & \small{1.52} & \small{(740, 2.65)} & \small{1.53} \\
\hline
\small{NGC 4945} & \small{(4.15, 2.30)} & \small{1.52} & \small{(3.45, 2.40)} & \small{1.07} \\
\hline
\small{NGC 1068} & \small{(16, 2.52)} & \small{0.65} & \small{(12, 2.50)} & \small{0.71} \\ 
\hline
\small{NGC 2146} & \small{(15, 2.50)} & \small{0.50} & \small{(12, 2.50)} & \small{0.52} \\
\hline
\small{ARP 299} & \small{(28, 2.15)} & \small{0.18}  & \small{(28, 2.05)} & \small{0.19} \\
\hline
\small{M31} & \small{(0.34, 2.40)} & \small{0.52}  &  \small{(0.19, 2.55)} & \small{0.46} \\
\hline
\small{M33} & \small{(0.44, 2.76)} & \small{0.44} & \small{(0.25, 2.89)} & \small{0.48}\\
\hline
\small{NGC 3424} & \small{(5.4, 2.22)} & \small{1.63} & \small{(5.4, 2.20)} & \small{1.51} \\
\hline
\small{NGC 2403} &\small{(0.75, 2.12)} & \small{0.38} & \small{(0.37, 2.15)} & \small{0.37} \\
\hline
\small{SMC} & \small{(0.038, 2.14)} & \small{1.90} & \small{(0.0149, 2.25)} & \small{5.50} \\
\hline
\small{Circinus} & \small{(6.6, 2.32)} & \small{0.92} &\small{(4.6, 2.30)} &\small{0.99}
\end{tabular}
\end{table}

\begin{figure}[t!]
    \centering
    \includegraphics[width=\columnwidth]{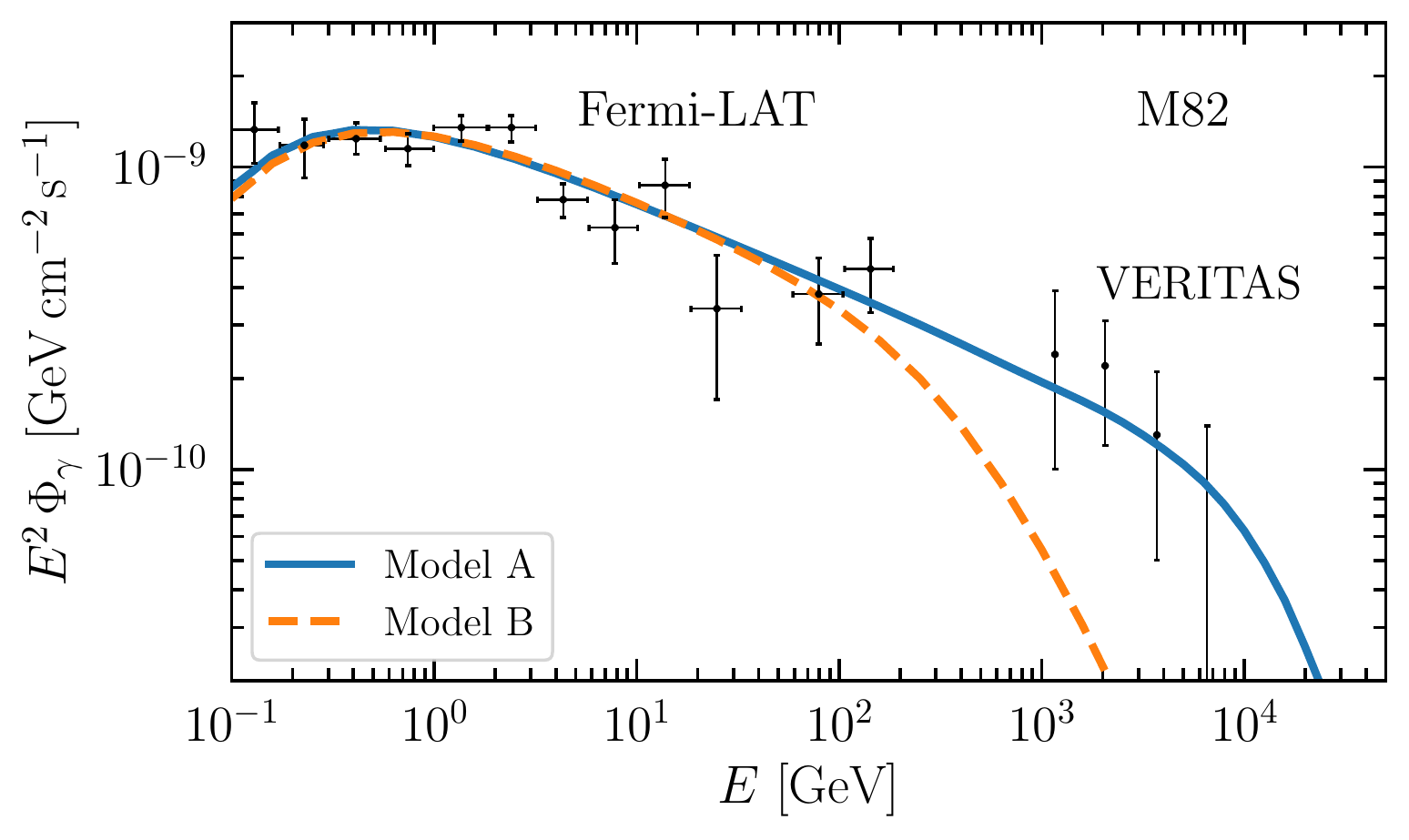}
    \caption{Best-fit SEDs for model A (dashed blu line) and model B (solid orange line) as inferred by Fermi-LAT data~\citet{Ajello:2020zna} and VERITAS data~\citep{2009Natur.462..770V}.\label{fig:m82}}
\end{figure}

The results of such an analysis are reported in Tab.~\ref{tab:data}. For most SBGs, current data are not able to statistically discriminate between the two models, although model A generally provides slightly smaller chi-squared values. On the other hand, for M82 and NGC 253 model B is disfavoured at ~$\sim2\sigma$ level. A tension at~$\sim 11\sigma$ with model B is found in the case of SMC, which however is mainly driven by low-energy data. Therefore, it is not determined by the high-energy production, and it may be influenced by the details of the leptonic production dominating at low energies. Thus, we do not consider this as a conclusive evidence in favor of model A. The importance of data above 100 GeV resides in the fact that the calorimetric fraction of model B rapidly drops to zero. This makes it highly difficult to accommodate high-energy data. In order to highlight this behaviour, in Fig.~\ref{fig:m82} we show the SED for M82 predicted by the two models in their best-fit scenario. This result is highly significant since, differently from~\cite{Krumholz:2019uom}, we find that model B is not able to accommodate high-energy data for this source.
This suggests that the diffusive model B cannot provide a full explanation of SFGs and SBGs emission above 1 TeV. On the other hand, it is worth noticing that the model B provides a higher calorimetric fraction for energies below $100 \ \text{GeV}$ with respect to model A. This implies that a smaller value of the SFR is required to fit the data. Therefore, sources like NGC 4945 get to be better fitted by this model because their emission can be better describe while satisfying the prior on the SFR deduced by IR and UV observations.

However, concerning the source NGC 4945, \cite{Ajello:2020zna} point out that the star-forming activity may be not responsible for the totality of its emissions due to the presence of an AGN activity. Indeed, model A cannot fully explain its gamma-ray data even with the highest value for the SFR ($4.15~{\rm M_\odot\, yr^{-1}}$) as allowed by the prior, leaving room for a possible AGN component. On the contrary, this is not the case for model B which requires $\dot{M}_* = 3.45~{\rm M_\odot\, yr^{-1}}$.

\begin{figure}[t!]
    \centering
    \includegraphics[width=\columnwidth]{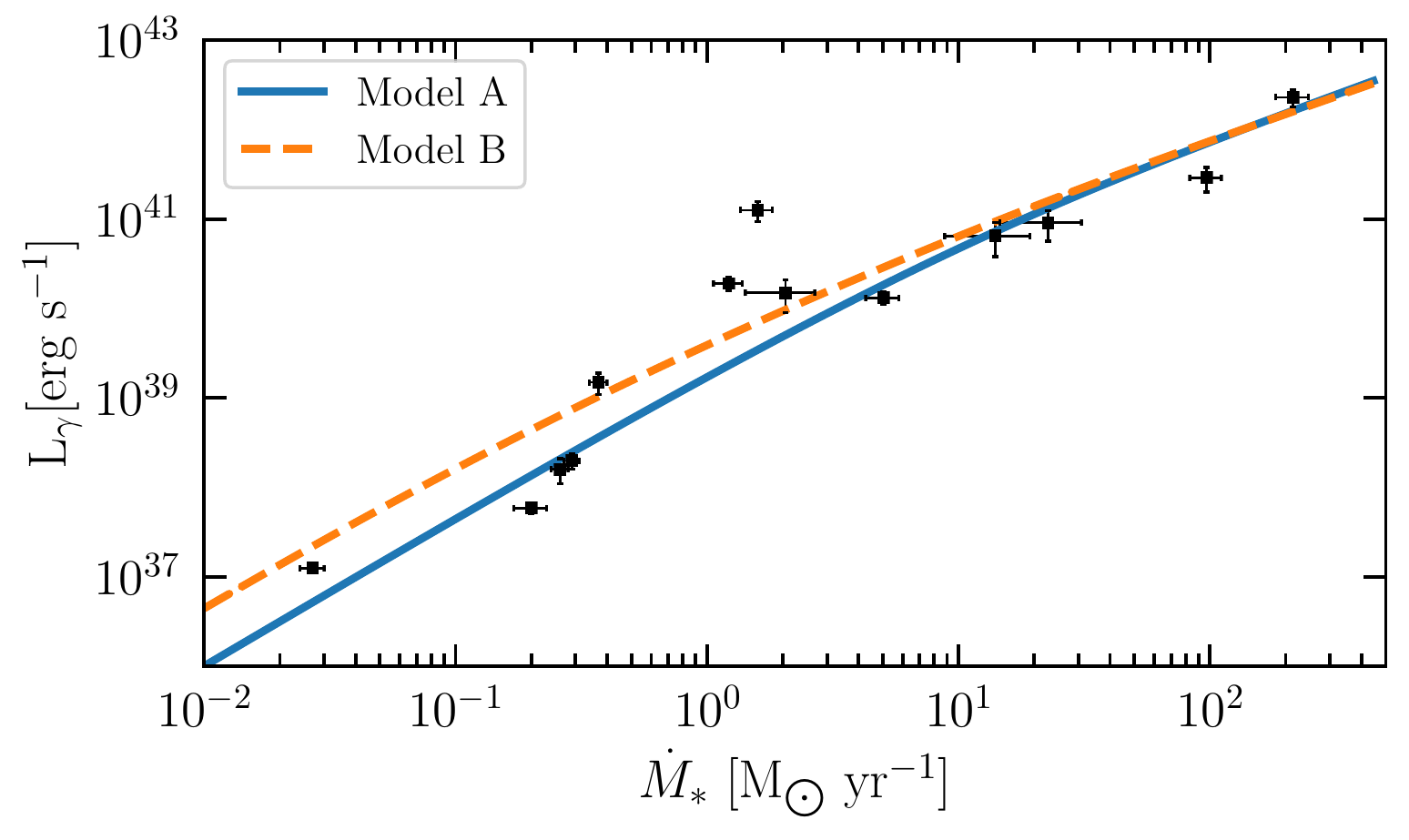}
    \caption{Comparison between the integrated luminosity between $(0.1-100) \ \text{GeV}$ for model A (solid blue line) and model B (dashed orange line), with the corresponding measurements of the 13 galaxies analysed~\citep{Kornecki:2020riv}.\label{fig:luminosita}}
\end{figure}

The different dependence on the SFR in the two models is highlighted in Fig.~\ref{fig:luminosita}, where we compare the gamma-ray luminosity $L_\gamma$ (integrated between 0.1 and 100 GeV) as a function of $\dot{M}_*$. In the plot, the integrated luminosity is computed assuming a reference value of the spectral index $(\Gamma = 2.2)$ for all the galaxies as in Refs.  \cite{Kornecki:2021xiy,Roth:2021lvk}. However, it is worth noticing that $L_\gamma$ depends only marginally on $\Gamma$. This allows for the comparison with the measurements of the individual sources even though the experimental SEDs are explained by different values for the spectral index (see Tab.~\ref{tab:data}). As can be clearly seen in the plot, at low SFR, the luminosity predicted by model B (dashed orange line) is higher than the one by model A. This just mirrors the higher calorimetric fraction of model B with respect to model A. On the contrary, for high values of the SFR, the two models predict the sources to be efficient calorimeters ($F_{\text{cal}} = 1$), leading to the same scaling of the luminosity with the SFR. 

The weak preference of current gamma-ray data for model A might already suggest the advection playing a key role as escape mechanism in SFGs and SBGs. However, more definitive conclusions might be only drawn with more gamma-ray data, especially at energies higher than 1~TeV. As will be discussed in the next subsection, the CR transport mechanism will be crucially probed by future gamma-ray telescopes.

\section{Forecast for the CTA telescope \label{sec:forecast}}

We perform a forecast analysis to quantitatively assess the ability of the future CTA telescope to discriminate between the two models of CR transport. In particular, we simulate future gamma-ray measurements using the public CTA information~\citep{CTAConsortium:2018tzg}. Motivated by the previous results, we assume the best-fit model A to generate CTA mock data and we determine the statistical power with which they favor model A over model B. We focus only on the local sources for which the expected SED is higher that the differential CTA sensitivity. According to model A which typically provides hard gamma-ray spectra, we expect that CTA telescope will observe at least four sources~\citep{Ambrosone:2021aaw}: SMC, M82, NGC~253 and Circinus galaxy. For each of these sources, we pursue the following procedure. Firstly, we bin the energetic range $(10^2 \ \text{GeV}- 10^5 \ \text{GeV})$, with the same binning provided by the CTA consortium. Secondly, for each source, we only consider the energetic bins for which the SED is higher than the CTA sensitivity. Hence, we calculate the expected number of signal events as
\begin{equation}
    \label{eq:signal_events}
    n_{\rm signal} = T_{\rm obs} \int_{\Delta E} A_{\rm eff}(E) \Phi_{\gamma, \rm A}(E) {\rm d}E
\end{equation}
where $\Phi_{\gamma, \rm A}$ is the gamma-ray flux predicted by the best-fit model A according to present data, $A_{\rm eff}$ is the CTA effective area, $\Delta E$ is the size of the energy bin and $T_{\rm obs} = 50~{\rm h}$ is the time of observation. We also take into account the number $n_{\text{bkg}}$ of background events associated to misidentified CRs. We remark that $n_{\text{bkg}}$ only depends on the declination of the source and the opening angle $\Delta \Omega$ of the observation. Considering that we expect gamma-rays mostly emitted by SBNs, we take $\Delta \Omega = {\rm max} [\Delta \Omega_{\rm res},  \Delta \Omega_{\rm SBN}]$ where $\Delta \Omega_{\rm res}$ is the CTA energy-dependent angular resolution function and $\Delta \Omega_{\rm SBN}$ represents the angular dimension of the source SBN. For all the sources except SMC (the nearest source) we have $\Delta \Omega_{\rm SBN} < \Delta \Omega_{\rm res}$. For SMC we consider the intrinsic extension of its SBN $\Delta \Omega_{\rm SBN} = 0.38^\circ$.

For each energy bin, we randomly generate $10^4$ numbers of events $n_{\rm obs}$ by means of a Poisson distribution with a mean value of $n_{\rm tot} = n_{\text{signal}}+ n_{\text{bkg}}$, namely $n_{\rm obs} \sim {\rm Poiss}(n_{\rm tot})$. Then, we estimate the empirical number of signal events simply as $\tilde{n}_{\rm signal} = n_{\rm obs}- n_{\text{background}}$. From this quantity we can calculate the empirical $\widetilde{\rm SED}$ assuming a generic power-law flux $E^{-2}$. In particular, we have
\begin{equation}
    \label{eq:emp_sed}
    \widetilde{\text{SED}}_{i} = \frac{\tilde{n}_{{\rm signal}, i}}{T_{\rm obs}  \int_{\Delta E} A_{\text{eff}}(E) \left(\frac{E}{1 \ \text{GeV}}\right)^{-2} {\rm d}E}
\end{equation}
where $i$ runs over energy bins. The reconstructed SEDs are affected by an uncertainty that can be directly estimated through the Poisson uncertainty on $n_{\rm obs}$ as
\begin{equation}\label{eq:exp_unc}
    \frac{\Delta \widetilde{\text{SED}}_{i}}{\widetilde{\text{SED}}_{i}} = \frac{\sqrt{n_{{\rm obs},i}}}{\tilde{n}_{\text{signal},i}}
\end{equation}
We emphasize that in this generation of mock data we simply consider the best-fit gamma-ray flux $\Phi_{\gamma, {\rm A}}$ provided by model A with current data. Hence, we neglect the intrinsic uncertainty on the expected gamma-ray flux as provided by the posterior distribution in Eq.~\eqref{eq:post}.We show the impact of the intrinsic uncertainty in the appendix. Accounting for the intrinsic uncertainty generally leads to larger Bayes factors and smaller p-values for SMC, M82, and NGC 253. Therefore, the results we obtain in the text are reasonably conservative.
\begin{figure*}[t!]
\centering
   \includegraphics[width=\columnwidth]{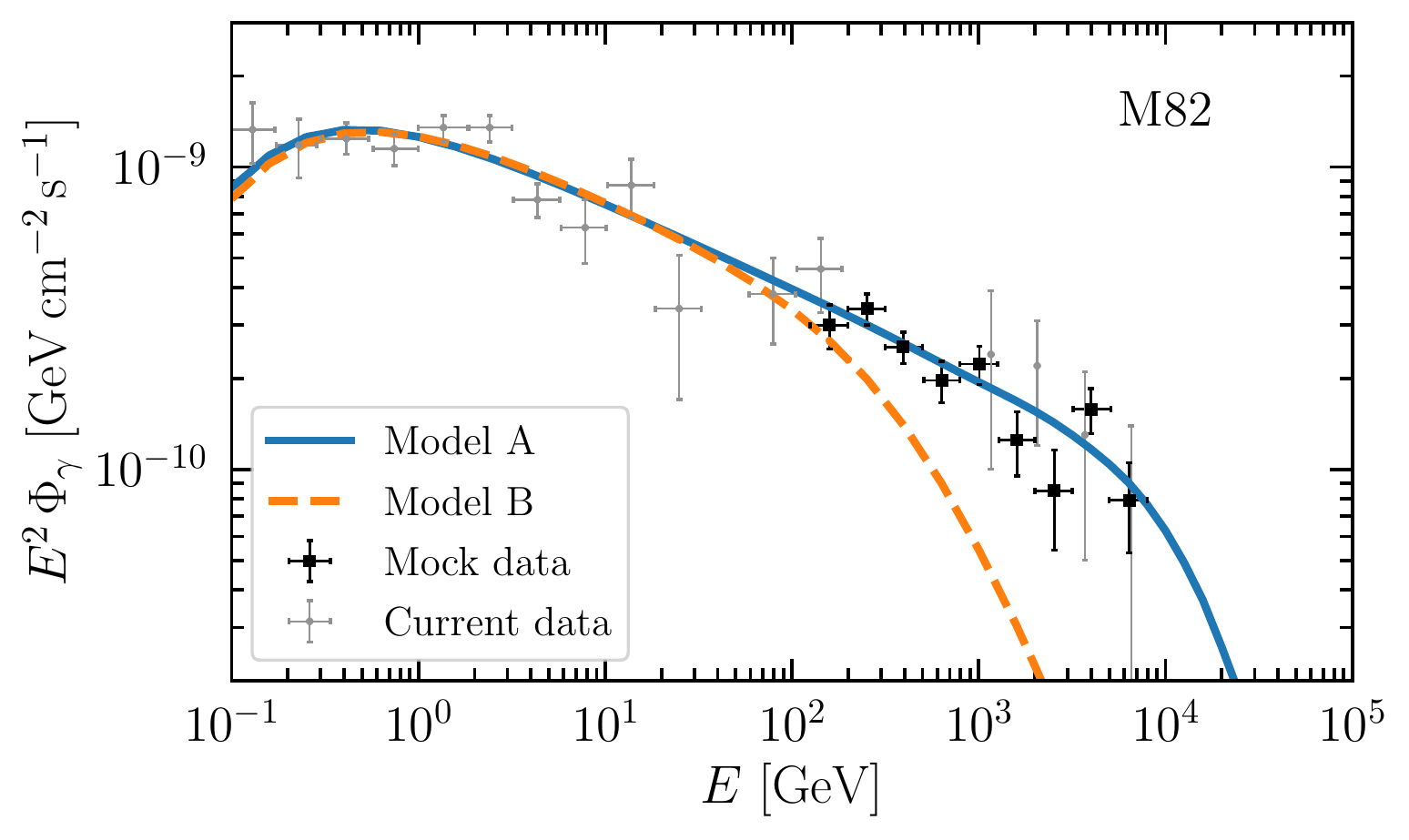}
   \includegraphics[width=\columnwidth]{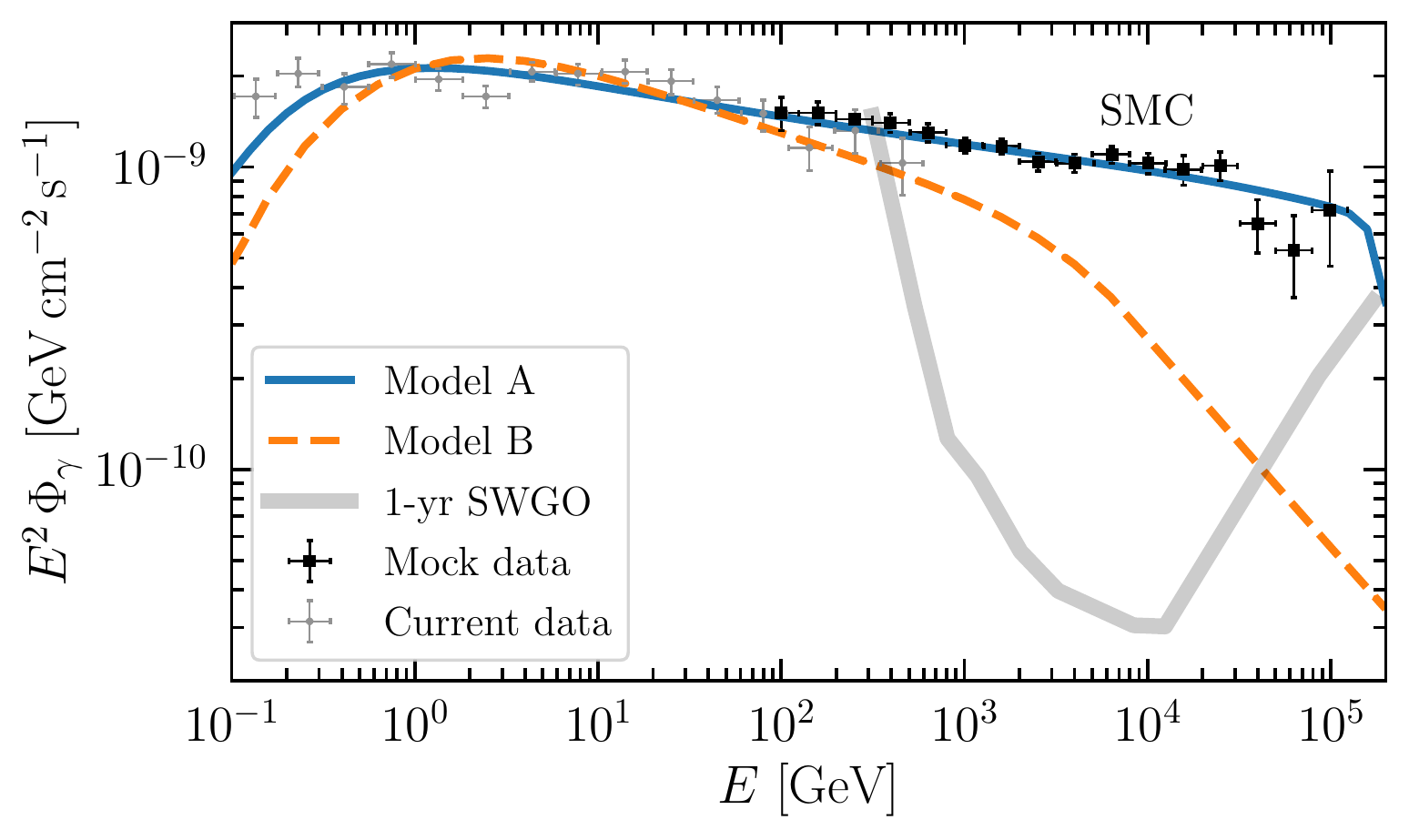}
   \includegraphics[width=\columnwidth]{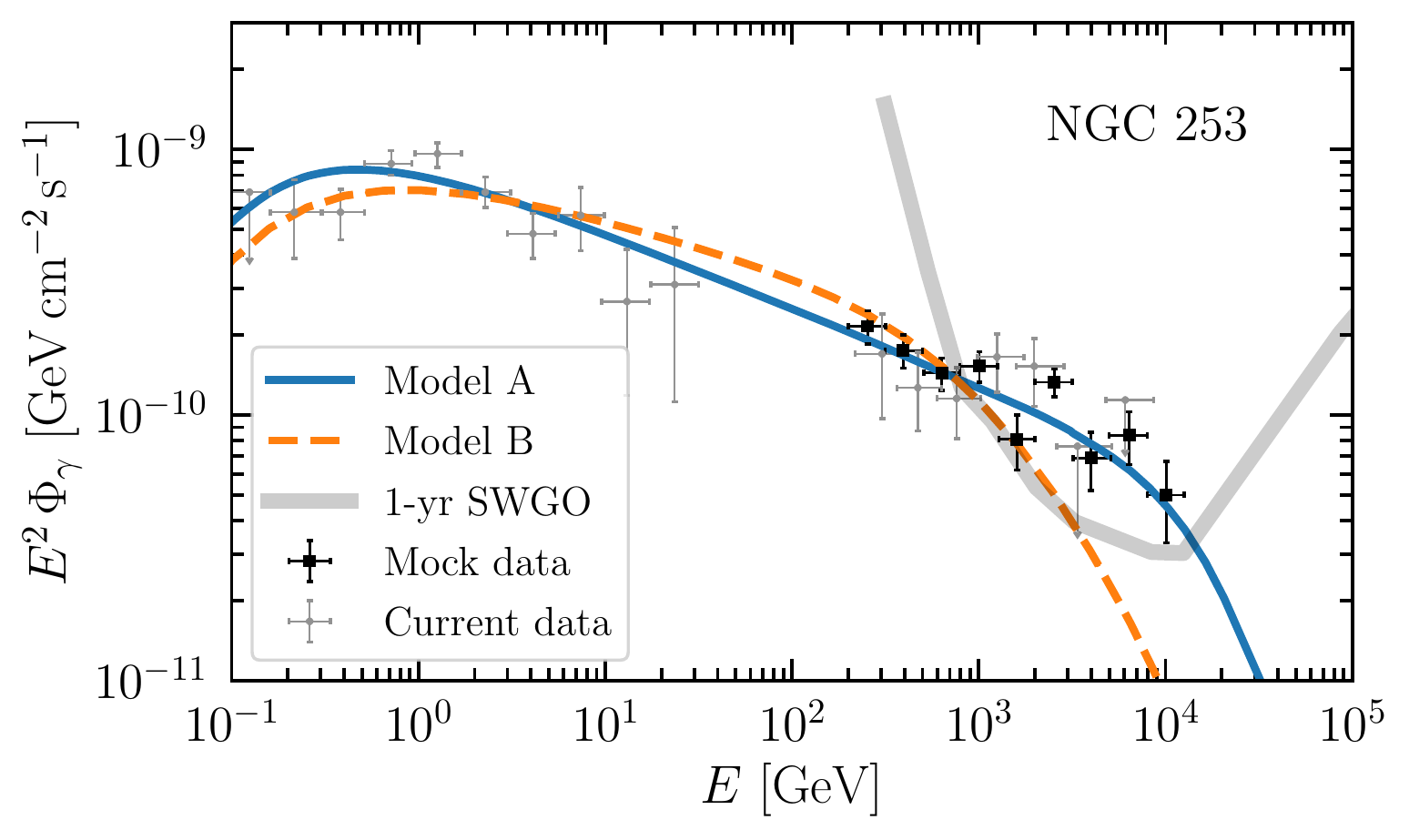}
   \includegraphics[width=\columnwidth]{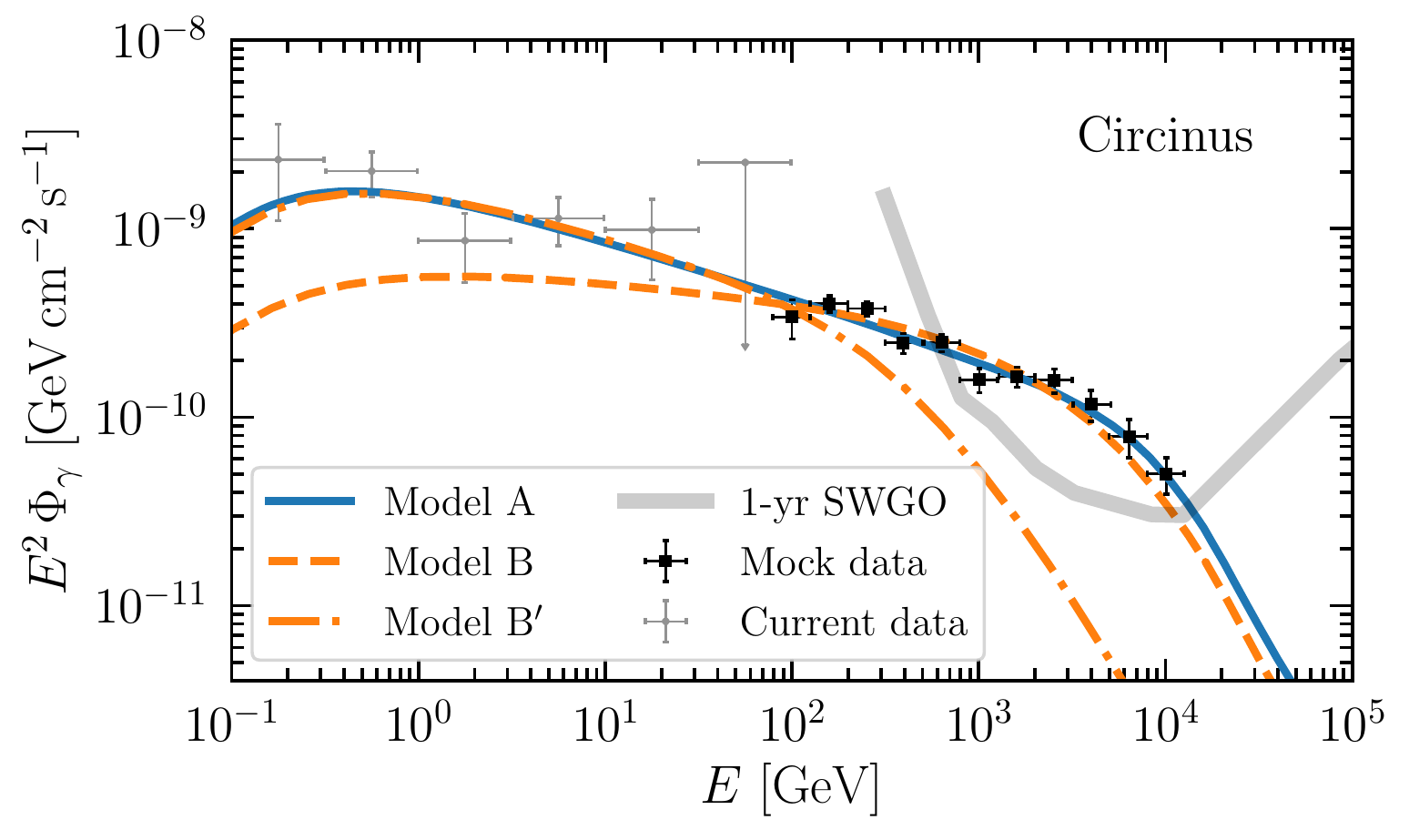}
   \caption{Spectral energy distributions of the most brightest sources compared to current (grey points) and one realization of CTA mock data (black points). The blue solid (orange dashed) lines show the best-fit scenario in case of model A (model B) of CR transport. For Circinus galaxy (bottom right panel), dashed-dotted orange line shows the best-fit SED for model B according to current data only. When present, grey curve represents the 1-year SWGO sensitivity to point-like sources in the Southern hemisphere~\citep{Hinton:2021um}.\label{fig:mockSEDs}}
\end{figure*}

For each of the $10^4$ mock data sets, we perform again the statistical analysis described in the previous section, including this time the mock data $\widetilde{\rm SED}$ as well. In Fig.~\ref{fig:mockSEDs} we show the updated best-fit scenarios for the four sources according to one realization of the CTA mock data. In particular, the solid blue (dashed orange) lines represent the best-fit SEDs according to model A (B). For M82, SMC, and NGC 253, they coincide with the one obtained with current data only. This stems from the fact that these sources have not only $\sim$GeV data which constrain their normalization (SFR), but also data above 100~GeV which constrain their shape ({\it i.e.} the spectral index). The only exception is Circinus Galaxy (bottom right panel) for which the current absence of data above 10 GeV leads to a different best-fit SED for model B (dot-dashed orange line referred to with ``model ${\rm B}^\prime$''). In the plots, we also show the expected $1 \ \text{yr}$ sensitivity of the SWGO experiment~\citep{Albert:2019afb,Hinton:2021um}. We emphasize that this future telescope is also expected to observe the brightest sources located in the Southern hemisphere in the $1-10$~TeV energy range.

We can then quantify the statistical confidence with which model B can be excluded by computing the $p$-value in a Frequentist approach as well as the Bayes factor $\mathcal{B}$ for the two models. The $p$-value is simply given by
\begin{equation}\label{eq:pvalue}
    p = \int_{\chi^{2}_{\text{min}}}^{\infty} f_k (x) \,{\rm d}x
\end{equation}
where $\chi^{2}_{\text{min}} = -2\ln\left[{\rm max}\,\mathcal{L}_{\rm B}(\dot{M}_*,\,\Gamma)\right]$ and $f_{k} (x)$ is the probability distribution function of a chi-square with $k$ degree of freedom. For each source, $k$ is given by total number of data points (current and mock data) minus two (number of free parameters). On the other hand, the Bayes factor is computed as
\begin{equation}
    \mathcal{B} = \frac{\int_{(\dot{M}_*,\,\Gamma)} \mathcal{L}_{\text{A}}({\rm SED} | \dot{M}_*,\,\Gamma) \ {\rm d}\dot{M}_* \ {\rm d}\Gamma}{\int_{(\dot{M}_*,\,\Gamma)} \mathcal{L}_{\text{B}}({\rm SED} | \dot{M}_*,\,\Gamma) \ {\rm d}\dot{M}_* \ {\rm d}\Gamma}
\end{equation}
where $\mathcal{L}_{\text{A}}$ and $\mathcal{L}_{\text{A}}$ are the likelihood functions for model A and B, respectively. We note that the two integrals are performed over the same two-dimensional phase space given by the uniform priors on $\dot{M}_*$ and $\Gamma$. For each source, we therefore obtain a distribution of expected $p$-values and Bayes factors given the different mock data sets.
\begin{table*}[t!]
\caption{\label{tab:mock} Results of the forecast analysis for the CTA telescope. For the four brightest sources, we report the mean $p$-values for model B and the mean Bayes factors for the two models corresponding to the distributions of $10^4$ CTA mock data sets, along with the ones deduced by current data. The columns 95\% and 68\% show the one-side intervals of expected $p$-values and Bayes factors corresponding to that confidence level.}
\centering          
\begin{tabular}{l | c c c c | c c c c }
\multirow{3}{*}{Source} & \multicolumn{4}{c|}{$p$-value} & \multicolumn{4}{c}{Bayes factor, $\mathcal{B}$} \\
& \multirow{2}{*}{Current data} & \multicolumn{3}{c|}{Mock data} & \multirow{2}{*}{Current data} & \multicolumn{3}{c}{Mock data} \\ \cline{3-5} \cline{7-9}
& & 95\% & 68\% & Mean & & 95\% & 68\% & Mean \\ \hline
SMC & $\ 4.3 \times 10^{-10}$ & $\ 9.1\times 10^{-33}$ & \ $4.4\times 10^{-35}$ & $\ 2.4 \times 10^{-36}$ & $\ 5.8\times 10^{10}$ & $\ 1.4 \times 10^{29}$ & $\ 6.6 \times 10^{30}$ & $\ 2.8\times 10^{31}$ \\ \hline
M82 & $2.3 \times 10^{-2}$ & $3.8\times 10^{-4}$ & $6.9\times 10^{-6}$ & $3.8\times 10^{-7}$ & $5.6 \times 10^{2}$ & $1.3 \times 10^3$ & $1.7 \times 10^6$ & $4.3 \times 10^7$ \\ \hline
NGC 253 & $1.5\times 10^{-2}$ & $4.2 \times 10^{-4}$ & $6.9\times 10^{-6}$ & $3.5 \times 10^{-6}$ & $2.5 \times 10^{2}$ & $3.4 \times 10^{5}$ & $4.9 \times 10^8$ & $\ 1.3 \times 10^{10}$ \\ \hline
Circinus & $4.1 \times 10^{-1}$ & $7.2 \times 10^{-2}$ & $1.3 \times 10^{-2}$ & $3.2 \times 10^{-3}$ & $1.0 $ & $8.3 \times 10^{1} $ & $2.5\times 10^{3}$ & $1.0\times 10^{4}$ \\ \hline
\end{tabular}
\end{table*}

In Tab.~\ref{tab:mock} we summarise the results of the above-describe forecast analysis. In particular, we report the mean values of the distributions of $p$-value and Bayes factor along with the corresponding values deduced with current data only. Moreover, we report the 68\% and 95\% one-sided intervals of the two distributions. As can be seen in the table, current data strongly disfavour model B for the SMC source. However, we find that with future CTA observations model B might be excluded at more than $2\sigma$ according to the expected small $p$-values. Furthermore, large Bayes factors are expected in favour of model A. This trend confirms that CTA will be able to firmly discriminate between the two models of CR transport within the SFGs and SBGs. 

\section{From Point Sources to Diffuse Fluxes\label{sec:diffuse}}

The observation of nearby galaxies provides valuable constraints on the parameters which define their point-like emissions. Equipped with such an information, we can now calculate the cumulative diffuse gamma-ray and neutrino fluxes correlated to unresolved SFGs and SBGs emission. As pointed out by~\cite{Roth:2021lvk}, model B~\citep{Krumholz:2019uom} is however inadequate at ultra-high energies, thereby making unreliable the comparison with IceCube neutrino observations. For this reason, hereafter we only consider model A.

We carry this calculation out by adopting a similar approach to the one described by~\cite{Ambrosone:2020evo}. The diffuse gamma-ray and neutrino flux can be obtained as
\begin{equation} 
\begin{aligned}\label{diffuse}
    \Phi^{\rm diff}_{\gamma,\nu} (E) = & \int_{0}^{4.2} \mathrm{d}z \int_{\dot{M}_{*,{\rm min}}}^{\infty} \mathrm{d}\log \dot{M}_* \, \frac{c \, d_c(z)^2}{H(z)} \\ 
    & \qquad \times \mathcal{S}_\mathrm{SFR}(z, \dot{M}_*) \, \Big\langle \Phi_{\gamma,\nu}\big(E,z| \dot{M}_*,\,\Gamma\big) \Big\rangle_\Gamma
\end{aligned}
\end{equation}
where we integrate over the whole SFGs and SBGs population in redshift $z$ and star formation rate $\dot{M}_*$, for which we consider the modified Schecter function $\mathcal{S}_\mathrm{SFR}(z,\dot{M}_*)$ reported by~\citet{Peretti:2019vsj}. Such a quantity has been obtained by fitting in the redshift interval $0 \leq z \leq 4.2$ the IR+UV data of a Herschel Source sample~\citep{Gruppioni:2013jna} after subtracting the AGN contamination~\citep{Delvecchio:2014dta}. For the cosmological Hubble parameter $H(z) = H_0 \sqrt{\Omega_M (1+z)^3 + \Omega_\Lambda}$ we take $H_0 = 67.74 \ \mathrm{km} \ \mathrm{s}^{-1}\mathrm{Mpc}^{-1} $, $\Omega_M = 0.31 $ and $\Omega_\Lambda = 0.69$, and $d_c(z)$ denotes the comoving distance. Finally, $\langle \Phi_{\nu,\gamma} \rangle_\Gamma$ is the emitted neutrino and gamma-ray fluxes averaged over the distribution of allowed spectral indexes. The main differences between the present calculation and the one in \citep{Ambrosone:2020evo} are threefold. Firstly, we take a different spectral index distribution, directly stemming from the $\Gamma$ values inferred by gamma-ray point-like data and reported in Tab.~\ref{tab:data}. This makes our estimates highly consistent with each other since the same CR transport model is used for point-like and diffuse analysis. In particular, we consider the spectral index distribution to be a Gaussian distribution with mean value of $4.36$ and standard deviation standard $0.2$. Secondly, we do not set by hand a lower limit for the SFR above which SBGs and SFGs are considered as good calorimeters. Indeed, as summarised by Eq.~\eqref{proton}, the sources which are dominated by either advection or diffusion losses naturally have their calorimetric fraction approaching zero, thus giving a negligible contribution to the diffuse fluxes. Therefore, we integrate from a minimum SFR $\dot{M}_{*,{\rm min}} = 0.038 \ \text{M}_{\bigodot} \text{yr}^{-1}$ according to our point-like analysis. Thirdly, we do not consider a free normalization for the diffuse flux, which is instead directly predicted under the aforementioned assumptions.
\begin{figure}[t!]
    \centering
    \includegraphics[width=\columnwidth]{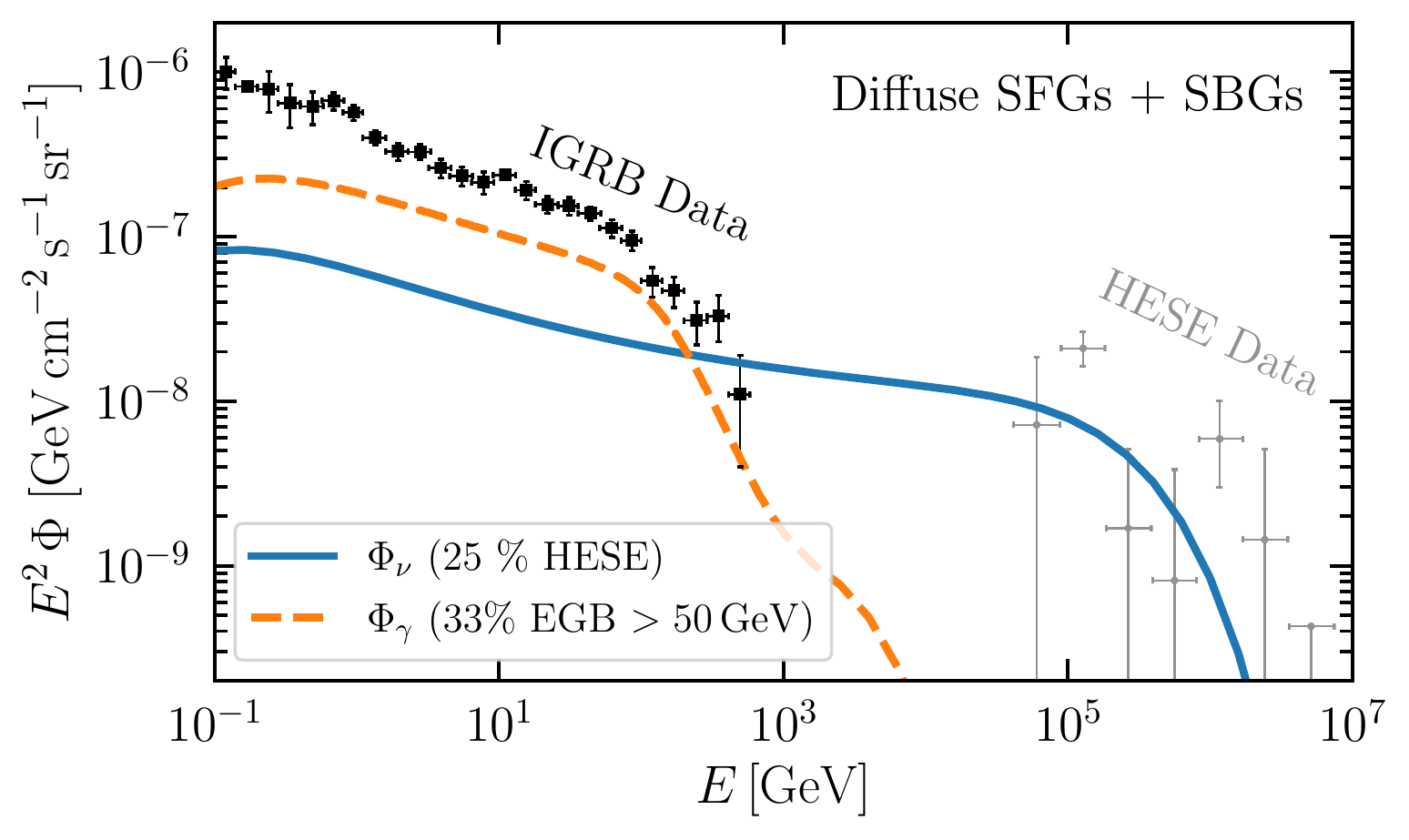}
    \caption{Diffuse neutrino (solid blue line) and gamma-ray (dashed orange line) flux from SFGs and SBGs predicted by model A for the CR transport. The contribution from electromagnetic cascades is also included. The data points correspond to the Fermi-LAT IGRB measurements~\citep{Fermi-LAT:2014ryh} and the IceCube HESE ones per neutrino flavour~\citep{IceCube:2020wum}.\label{fig:diffuseflux}}
\end{figure}

Fig.~\ref{fig:diffuseflux} shows the diffuse gamma-ray and neutrino fluxes compared with the Isotropic Gamma-Ray Background (IGRB)~\citep{Fermi-LAT:2014ryh} and IceCube HESE~\citep{IceCube:2020wum} data, respectively. The gamma-ray flux takes also into account the electromagnetic cascade contribution, calculated with the $\gamma$-Cascade public code \citep{Blanco:2018bbf}. We find a gamma-ray flux similar to previous estimates~\citep{Peretti:2019vsj,Ambrosone:2020evo,Owen:2021qul}. In particular, we predict the SFGs and SBGs to provide an important contribution to the IGRB. Nevertheless, differently from the results of Refs.~\citep{Tamborra:2014xia,Roth:2021lvk}, our prediction is consistent within $1\sigma$ uncertainties given with the non-blazar limits \citep{TheFermi-LAT:2015ykq,Lisanti:2016jub,Bechtol:2015uqb,Yoshida:2020div}, since it corresponds to $33\%$ of the total EGB integrated above 50 GeV. This is highly significant since our result is obtained without any fine-tuning of parameters. Furthermore, as far as the neutrino production is concerned, we predict a flux which can explain a considerable fraction of the IceCube observations. Through the IceCube effect area, we find that our model produces 25 events with energy higher than 30~TeV after 7.5~years of data-taking. This corresponds to roughly $25\%$ of the totality of the HESE observed by the IceCube collaboration. Hence, differently from~\cite{Roth:2021lvk}, we find a significantly higher neutrino contribution without violating the diffuse gamma-ray constraints. We point out that not only model A seems to be preferred by current gamma-ray point-like data, but it may also provide a better explanation of the diffuse gamma-ray and neutrino observations in a multi-messenger context without exceeding the extra-galactic limit on non-blazar sources.

\section{Conclusions\label{sec:conclusion}}

In the present paper, we have investigated the phenomenological consequences of two different models (model A~\citep{Peretti:2018tmo} and model B~\citep{Krumholz:2019uom}) for cosmic-ray transport within the cores of starforming and starburst galaxies. We have shown that current point-like observations by Fermi-LAT, VERITAS and H.E.S.S. gamma-ray telescopes already prefer model A, especially for the brightest sources SMC, M82 and NGC 253. Then, we have performed a forecast analysis for the CTA telescope which will potentially observe the gamma-ray emission from nearby galaxies at higher energies. Interestingly, we have found that future CTA observations have the potential to firmly discriminate between model A and model B, providing the latter a highly-suppressed emission above a few TeV. We emphasize that in the Southern hemisphere a crucial role will be also played by the upcoming SWGO telescope thanks to a larger field of view and a longer duty-cycle compared to CTA. The comparison between the two models depends crucially on the hadronic production above~$\sim 1$~TeV, which is present in model A and suppressed in model B due to diffusion. Since this hadronic production generally follows a power-law spectrum, the results we obtain are roughly independent of the details of the models (e.g. choice of the astrophysical parameters), and are mainly determined by the spectral index, normalization, and maximal cosmic-ray energy, which we obtain from current gamma-ray data on star-forming galaxies. Finally, we have employed the information inferred by the local gamma-ray observations to consistently and robustly estimate the diffuse gamma-ray and neutrinos emission from the whole population of starforming and starburst galaxies. We have found that model A predicts a $25\%$ contribution of these sources to the IceCube HESE data, in agreement with the gamma-ray limits on the non-blazar component.
This result confirms that cosmic reservoirs are of paramount importance for the description of high-energy neutrino observations, even though they cannot explain the whole astrophysical flux observed.

\begin{acknowledgements}
    This work was partially supported by the research grant number 2017W4HA7S ``NAT-NET: Neutrino and Astroparticle Theory Network'' under the program PRIN 2017 funded by the Italian Ministero dell'Universit\`a e della Ricerca (MUR). The authors also acknowledge the support by the research project TAsP (Theoretical Astroparticle Physics) funded by the Istituto Nazionale di Fisica Nucleare (INFN). The work of DFGF is partially supported by the {\sc Villum Fonden} under project no.~29388. This project has received funding from the European Union's Horizon 2020 research and innovation program under the Marie Sklodowska-Curie grant agreement No.~847523 ‘INTERACTIONS’. 
\end{acknowledgements}

%
%

\bibliographystyle{aa}
\bibliography{references}

\begin{appendix}

\section{Impact of source uncertainties on mock data generation\label{app:1}}

We here investigate how the mock data generation for the CTA telescope is affected by the current uncertainties on the source parameters. Such uncertainties are simply defined by the posterior distribution in Eq.~\eqref{eq:post} that is obtained with current gamma-ray data. Differently from the analysis presented in Sec.~\ref{sec:forecast}, for each source we produce mock data by sampling the star formation rate and the spectral index from the posterior distribution of model A, namely $(\dot{M}_*,\,\Gamma) \sim p_{\rm A}(\dot{M}_*,\,\Gamma|{\rm SED})$. These parameters determine the gamma-ray flux $\Phi_{\gamma, {\rm A}}$ which is then employed to compute the expected number of signal events $n_{\rm signal}$ according to Eq.~\eqref{eq:signal_events}. This time, we consider the observed number of events simply as $n_{\rm obs} = n_{\rm signal} + n_{\rm bkg}$ and estimate the empirical $\widetilde{\rm SED}$ directly from $n_{\rm signal}$ (see Eq.~\eqref{eq:emp_sed}). As before, the uncertainty on the reconstructed SEDs is deduced from the Poisson uncertainty on the mock measurements $n_{\rm obs}$. In this way, for each source we produce $\mathcal{O}(10^4)$ mock data sets which are analyzed assuming model A and model B. As described in Sec.~\ref{sec:forecast}, we compute the $p$-values to test model B and the Bayes factors to compare the two models.

\begin{figure}[t!]
   \centering
   \includegraphics[width=\columnwidth]{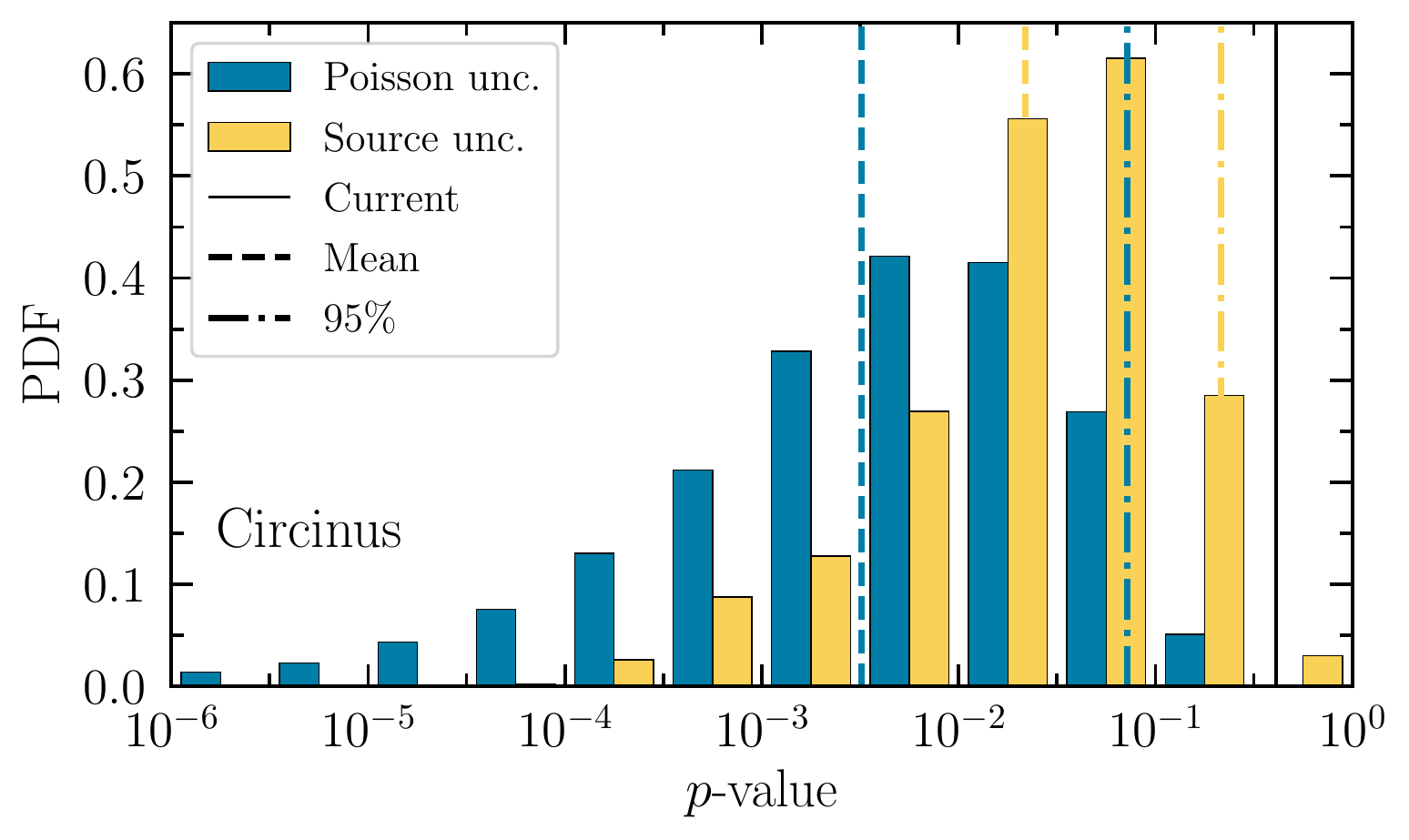}
   \includegraphics[width=\columnwidth]{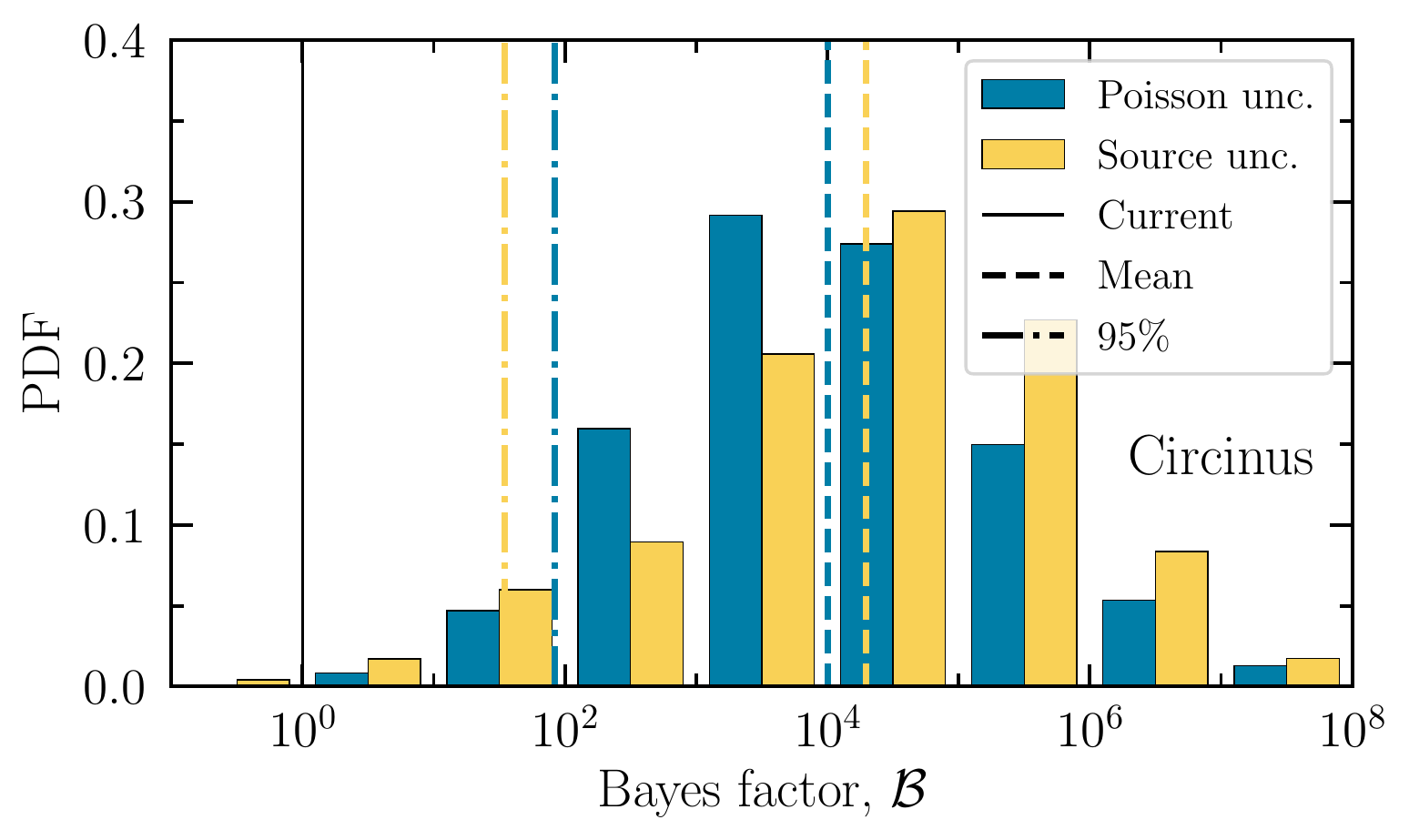}
   \caption{Distributions of $p$-values (top panel) and Bayes factors (bottom panel) obtained by generating CTA mock data according to a Poisson distribution (blue color) and to the posterior distribution of source parameters (yellow color). The vertical lines show both the current value (solid lines) with collected data, and the mean (dotted lines) and 95\% (dot-dashed) values corresponding to the mock distributions.\label{fig:circinus_dist}}
\end{figure}

The results of this analysis are reported in Tab.~\ref{tab:sampling}.These results are quite more stringent than the one given by the previous case because for M82, NGC 253 and SMC the variability of the fit given by model A is pretty small and therefore the prediction of this model are quite in disagreement with the one predicted by model B. In particular, for M82 and NGC 253, the mean Bayes factors are respectively of the order of $10^{23}$ and $10^{9}$, which are much greater than the values obtained through the Poisson generation in Sec~\ref{sec:forecast}. Correspondingly, the rejection p-values are lower than the Poissonian mock data. In fact, for SMC, the average p-value is given by $2.0 \times 10^{-38}$, which is 2 order of magnitudes lower than the one provided by Poissonian mock data. The only exception is Circinus for which we just, currently, have six data points making its posterior distribution quite unconstrained. In this case, we obtain similar results for the two different approaches to mock data generation. We show this in Fig.~\ref{fig:circinus_dist} where in the top (bottom) panel we compare the distributions of for $p$-values (Bayes factors) for the two mock datasets. We also highlight the mean and the $95\%$ C.L. values with dashed and dot-dashed lines, respectively, as well as the current values with collected gamma-ray data (solid lines). Even though the width of the distributions for the Poisson uncertainty and the source uncertainty are similar in this case, the distribution for latter case is peaked at higher values of Bayes factor. Therefore, it is conservative to consider only the Poisson uncertainty, as we do in the main text.

\begin{table}[tbh!]
\caption{\label{tab:sampling} Results of the CTA forecast analysis once the mock data are generated according the source posterior distribution in Eq.~\eqref{eq:post} obtained with current data. As in Tab.~\ref{tab:mock}, we report the 95\%, 68\% and mean $p$-values testing model B as well as the Bayes factors comparing model A with model B.}
\centering          
\begin{tabular}{c l | c c c c}
\multicolumn{2}{c|}{\multirow{2}{*}{\small{Mock data}}} & \multicolumn{4}{c}{Source} \\
& & SMC & M82 & NGC 253 & Circinus \\ \hline
\multicolumn{1}{c|}{\multirow{3}{*}{\rotatebox[origin=c]{90}{$p$-value}}} & 95\% & \small{$2.0 \times 10^{-28}$} & \small{$2.8\times 10^{-5}$} & \small{$8.0\times 10^{-9}$} & \small{$2.2\times 10^{-1}$} \\
\multicolumn{1}{c|}{} & 68\% & \small{$1.7\times 10^{-34}$} & \small{$9.4\times 10^{-6}$} & \small{$2.5\times 10^{-9}$} & \small{$4.8\times10^{-2}$} \\
\multicolumn{1}{c|}{} & \small{Mean} & \small{$1.0\times 10^{-38}$} & \small{$1.8\times 10^{-6}$} & \small{$5.7\times 10^{-10}$} & \small{$2.2\times 10^{-2}$} \\ \hline \hline 
\multicolumn{1}{c|}{\multirow{3}{*}{\rotatebox[origin=c]{90}{$\mathcal{B}$}}} & 95\% & \small{$1.7\times 10^{29}$} & \small{$1.8\times 10^{19}$} & \small{$1.4\times 10^{6}$} & \small{$3.5\times 10^{1}$} \\
\multicolumn{1}{c|}{} & 68\% & \small{$1.2\times 10^{31}$} & \small{$1.2\times 10^{21}$} & \small{$1.1\times10^{8}$} & \small{$6.0\times 10^{3}$} \\
\multicolumn{1}{c|}{} & \small{Mean} & \small{$2.4\times 10^{33}$} & \small{$1.6\times 10^{23}$} & \small{$3.8\times 10^{9}$} & \small{$2.0\times10^{4}$}
\end{tabular}
\end{table}

\end{appendix}
\end{document}